\documentclass[aps, a4paper, superscriptaddress, nofootinbib, twocolumn, 10pt]{revtex4}
\usepackage{amsmath,amssymb, bm}
\usepackage{dsfont}
\usepackage{epsfig}
\usepackage{graphicx}
\usepackage{slashed}
\usepackage{color}
\usepackage{subfigure}
\usepackage{hyperref}
\newcommand{\be}{\begin{equation}}
\newcommand{\ee}{\end{equation}}
\newcommand{\wt}{\widetilde}
\newcommand{\beq}{\begin{equation}}
\newcommand{\eeq}{\end{equation}}
\newcommand{\bea}{\begin{eqnarray}}
\newcommand{\eea}{\end{eqnarray}}
\newcommand{\ba}{\begin{align}}
\newcommand{\ea}{\end{align}}
\newcommand{\bfig}{\begin{figure}}
\newcommand{\efig}{\end{figure}}

\newcommand{\wh}{\widehat}

\begin{document}
~\vspace{0.5cm}
\title{ Conformal mapping of the  Borel  plane: going beyond perturbative QCD  }
\author{Irinel Caprini}
\affiliation{Horia Hulubei National Institute for Physics and Nuclear Engineering, P.O.B. MG-6, 077125 Bucharest-Magurele, Romania}

\begin{abstract}  The power corrections in the Operator Product Expansion (OPE) of QCD correlators can be viewed mathematically as an illustration of the transseries concept, which allows to recover a function from its asymptotic divergent expansion. Alternatively, starting from the divergent behavior of the perturbative QCD encoded in the singularities in the Borel plane, a modified expansion can be defined by means of the conformal mapping of this plane. A comparison of the two approaches concerning  their ability to recover nonperturbative properties of the true correlator was not explored up to now.  In the present paper, we make a first attempt to investigate this problem. We use for illustration the Adler function and observables expressed as integrals of this function along contours in the complex energy plane.  We show that the expansions based on the conformal mapping of the Borel plane  go beyond finite-order perturbation theory, containing an infinite number of terms when reexpanded in powers of the coupling. Moreover, the expansion functions exhibit nonperturbative features of the true function,  while the expansions have a tamed behavior at large orders and are expected even to be convergent. Using these properties, we argue that there are no mathematical reasons for supplementing the expansions based on the conformal mapping of the Borel plane  by additional arbitrary power corrections. Therefore, we make the conjecture that they provide  an alternative to the standard OPE in approximating the QCD correlator. This conjecture allows to slightly improve the accuracy  of the strong coupling extracted from the hadronic $\tau$ decay width.  Using the optimal expansions based on conformal mapping and the contour-improved prescription of renormalization-group resummation, we obtain  $\alpha_s(m_\tau^2)=0.314 \pm 0.006$, which implies $\alpha_s(m_Z^2)=0.1179 \pm 0.0008$.

\end{abstract}

\maketitle
\vspace{0.2cm} 
\section{Introduction}
Perturbation theory is known to lead to divergent series for many quantities in Quantum Mechanics and  Quantum Field Theory (QFT). This surprising fact was first noticed in 1952 by Freeman Dyson \cite{Dyson:1952tj}, who argued that the perturbation expansions in QED cannot be convergent since the expanded functions are singular at the expansion point. This discovery   set a challenge for a radical reformulation of perturbation theory (PT). 
To give the divergent series a precise meaning, Dyson proposed to interpret it as asymptotic to  the exact function, which changed the entire philosophy of perturbation theory. Perturbation theory yields, at least in principle, the values of all  the perturbative coefficients. This can tell us whether the series is convergent or not. But what we want to know is under what conditions the expanded function can be recovered.  If the series were  convergent, the knowledge of all the perturbative coefficients  would uniquely determine the function. On the other hand, there are infinitely many functions having the same  asymptotic expansion.

A divergent power series indicates that the expanded function is singular at the expansion point. This means that the Green functions in QFT are expected to be singular at the origin of the  coupling  plane. In the case of QED, the singular behavior was discovered by Dyson through his original reasoning  \cite{Dyson:1952tj}. For QCD, the existence of the singularity at zero coupling was demonstrated by 't Hooft \cite{tHooft}, using unitarity, analyticity and renormalization group invariance. The divergence can be inferred alternatively from particular classes of Feynman diagrams, which indicate a factorial growth of the expansion coefficients in both QED \cite{Lautrup:1977hs, Broadhurst:1992si} and QCD \cite{Beneke:1994qe, Beneke:1992ch, Beneke:1998ui}.   Compelling evidence for this behavior is provided also by lattice calculations \cite{Bauer:2011ws}.

Borel summation is known to be a useful tool for dealing with divergent series. The large-order properties of the expansion coefficients of a function are encoded in the singularities of its Borel transform in the Borel plane. These singularities (in particular the infrared (IR) renormalons produced by the low momenta in the Feynman diagrams) induce ambiguities in the Laplace-Borel integral by which the original function is recovered from its Borel transform. According to the standard view, this  indicates that perturbation theory is not complete and must be supplemented by  nonperturbative terms in order to recover the true function  \cite{Mueller1985, Mueller:1993pa, Beneke:1998ui}. In QCD, these terms,  exponentially small in the strong coupling, are identified with the power corrections in the Operator Product Expansion (OPE) of the Green functions \cite{Shifman:1978bx}. 

 In mathematical terms, in the so-called hyperasymptotic theory, the power corrections can be interpreted as a first piece of a transseries, i.e.,  a sequence of truncated series, each of them  exponentially small in the expansion parameter of the previous one,  which allow to recover the expanded function from its asymptotic divergent expansion  (see \cite{BerryHowls, Howls, Dorigoni:2014hea} and references therein). The hyperasymptotic approximation  has been used  in QCD  in order to separate the truncated perturbative series from the nonperturbative terms in the calculation of several observables \cite{Ayala:2019uaw, Ayala:2019hkn}.

On the other hand, a reformulation of perturbative QCD has been defined recently  using the method of conformal mapping for ``series acceleration'', i.e., for enlarging the domain of convergence  of power series  and for increasing their  rate of convergence. The conformal mappings have been applied   a long time ago  to the scattering amplitudes in particle physics \cite{CiFi, Frazer, CiCiFi}, and more recently to the perturbative expansions in QFT \cite{Seznec:1979ev, ZinnJustin:2010ng}. In particular, as shown in \cite{Mueller:1993pa, Altarelli:1994vz}, the spurious power corrections  in the QCD correlators, which are due to the large momenta in the  Feynman integrals and are formally related to the ultraviolet (UV) renormalons, can be removed by means of a conformal mapping of the Borel plane.  However,  the conformal mapping used in  \cite{Mueller:1993pa, Altarelli:1994vz} does not ensure the best convergence rate of the corresponding series. As proved  in  \cite{Caprini:1998wg}, an optimal conformal mapping can be defined, which achieves the analytic continuation of the Borel transform in the whole Borel plane and has the best  asymptotic convergence rate. The properties of the perturbative expansions in QCD improved by means of this mapping have been investigated in \cite{Caprini:2000js, Caprini:2001mn}, and   the method has been further considered in   \cite{Cvetic:2001sn, Jeong:2002ph, Caprini:2009vf, Caprini:2011ya, Abbas:2012fi,Abbas:2013usa, Caprini:2018agy, Caprini:2019kwp} (see also the reviews \cite{Caprini:2017ikn, Caprini:2019osi}).

As shown in  \cite{Caprini:1998wg}, the optimal conformal mapping of the Borel plane for QCD incorporates information on the position of the IR and UV renormalons.  On the other hand, the power corrections are introduced in the standard OPE precisely to take into account the effect of the IR renormalons.  This implies,  as remarked in \cite{Caprini:2019osi}, that the method of conformal mapping can be viewed as an alternative to the transseries approach.   In the present paper we discuss in more detail this problem and argue that  the method of conformal mapping provides a systematic representation which allows to recapture nonperturbative features of the exact function, without the need for additional power corrections.  We note that the same problem was discussed recently in the mathematical literature \cite{Costin:2017ziv,  Costin:2019xql, Florio:2019hzn}, where the possibility of recovering the exact function from the coefficients of its asymptotic  perturbative expansion was demonstrated in several cases where the exact function is known.

The outline of the paper is as follows: in the next section we briefly review the perturbative expansion of the Adler function for massless quarks and in Sec. \ref{sec:conf} we define a reformulation of perturbation theory for this function using the conformal mapping of the Borel plane.   Section \ref{sec:resurg} contains our arguments in favour of the idea that the perturbative expansions based on the optimal conformal mapping of the Borel plane represent an alternative to the transseries. In Sec.  \ref{sec:moments} we discuss the perturbative expansions  of the moments of the spectral function, using recent results on their singularities  in the Borel plane \cite{Caprini:2019kwp, Boito:2020hvu}. In Sec. \ref{sec:delta0},  we consider in particular the  contour-improved (CI) and  fixed-order (FO)  expansions of the $\tau$ hadronic width and in Sec. \ref{sec:alphas} we present a new determination of the strong coupling $\alpha_s$ from $\tau$ hadronic width.   Finally, section \ref{sec:conc} contains our conclusions.

\section{Adler function in perturbative QCD}\label{sec:Adler}
 We  consider the reduced Adler function \cite{Beneke:2008ad}
\beq\label{eq:D}
\widehat{D}(s) \equiv 4 \pi^2 D(s) -1,
\eeq
where $D(s)=-s \,d\Pi(s)/ds$ is the logarithmic derivative of the invariant amplitude $\Pi(s)$ of the two-current correlation tensor.  From general principles of field theory, it is known that $\wh D(s)$ is an analytic function of real type (i.e., it satisfies the  Schwarz reflection property $\wh D(s^*)=\wh D^*(s)$) in the complex $s$ plane cut along the timelike axis for $s\ge 4 m_\pi^2$. 

 In QCD perturbation theory,  $\wh D(s)$ is expressed as an expansion  
\beq\label{eq:hatD}
\widehat{D}(s) =\sum\limits_{n\ge 1} [a(\mu^2)]^n \,
\sum\limits_{k=1}^{n} k\, c_{n,k}\, (\ln (-s/\mu^2))^{k-1},
\eeq
in powers of the renormalized strong coupling $a(\mu^2) \equiv \alpha_s(\mu^2)/\pi$, defined in a certain renormalization scheme (RS) at the renormalization scale $\mu$. Since the series is divergent, the representation is actually symbolic and has to be given a meaning.

The  coefficients  $c_{n,1}$ in (\ref{eq:hatD}) are obtained from the calculation of  Feynman diagrams, while  $c_{n,k}$ with $k>1$ are expressed in terms of  $c_{m,1}$ with $m< n$  and the perturbative coefficients $\beta_n$ of the $\beta$ function, which governs the variation of the QCD coupling with the scale $\mu$ in each RS:
 \begin{equation}\label{eq:rge}
 -\mu\frac {d a_\mu}
{d\mu}\equiv \beta(a_\mu)=\sum_{n\ge 1}
\beta_n a_\mu^{n+1}. \end{equation}
In $\overline{{\rm MS}}$ scheme, the coefficients $\beta_n$ have been calculated to five loops  (see \cite{Baikov:2016tgj} and references therein). The first two coefficients do not depend on the RS and are expressed in terms of the number $n_f$ of active flavours as:
\beq\label{eq:betai}
\beta_1=\frac{11}{2}-\frac{1}{3}n_f,\quad \beta_2=\frac{51}{4}-\frac{19}{12}n_f.
\eeq 

For a large spacelike value $s<0$, one can choose in (\ref{eq:hatD}) the scale $\mu^2=-s$, and obtain the renormalization-group improved expansion
\beq\label{eq:hatD1}
\widehat{D}(s) =\sum\limits_{n\ge 1} c_{n,1}\, [a(-s)]^n,
\eeq
where $a(-s)\equiv \alpha_s(-s)/\pi$ is the running coupling. The expansions (\ref{eq:hatD})  and (\ref{eq:hatD1}) are often used also for complex values of $s$ plane, outside the timelike axis $s>0$.

The Adler function  was calculated in the $\overline{{\rm MS}}$ scheme to order $\alpha_s^4$  (see \cite{Baikov:2008jh} and references therein). For $n_f=3$, the leading coefficients $c_{n,1}$  have the values:
\be\label{eq:cn1}
c_{1,1}=1,\,\, c_{2,1}=1.640,\,\, c_{3,1}=6.371,\,\, c_{4,1}=49.076.
\ee
Estimates of the next coefficient $c_{5,1}$ have been made in several papers (see \cite{Boito:2018rwt, Caprini:2019kwp} and references therein). We shall use in our analysis the range
\be\label{eq:c51}
c_{5,1}=277\pm 51,
\ee
derived recently in \cite{Boito:2018rwt}.  

At high orders $n$, the coefficients increase factorially, more exactly $c_{n,1}\approx K\,b^n n !\, n^{c}$,  where  $K$, $b$ and $c$ are  constants  \cite{Beneke:1998ui}.  Therefore, the series (\ref{eq:hatD})  has zero radius of convergence and can be interpreted only as an asymptotic expansion to $\widehat{D}(s)$ for $a(\mu^2)\to 0$. This indicates the fact that the Adler function, viewed as a function of the strong coupling $a$, is singular at the origin $a=0$ of the coupling plane. Actually, as shown by 't Hooft \cite{tHooft}, the function  $\widehat{D}$  is  analytic only in a horn-shaped region in the half-plane $\text{Re}\, a>0$, of  zero opening angle near $a=0$. 

In some cases, the expanded functions can be recovered  from their divergent expansions through Borel summation. The Borel transform of the Adler function is defined by the power series
\be\label{eq:B}
 B_{\widehat D}(u)= \sum_{n=0}^\infty  b_n\, u^n,
\ee
where the coefficients $b_n$ are related to the perturbative coefficients $c_{n,1}$ by 
\be\label{eq:bn}
 b_n= \frac{c_{n+1,1}}{\beta_0^n \,n!}\,.
\ee
Here we used the standard notation $\beta_0=\beta_1/2$.

\begin{figure}
\includegraphics[width=6.6cm]{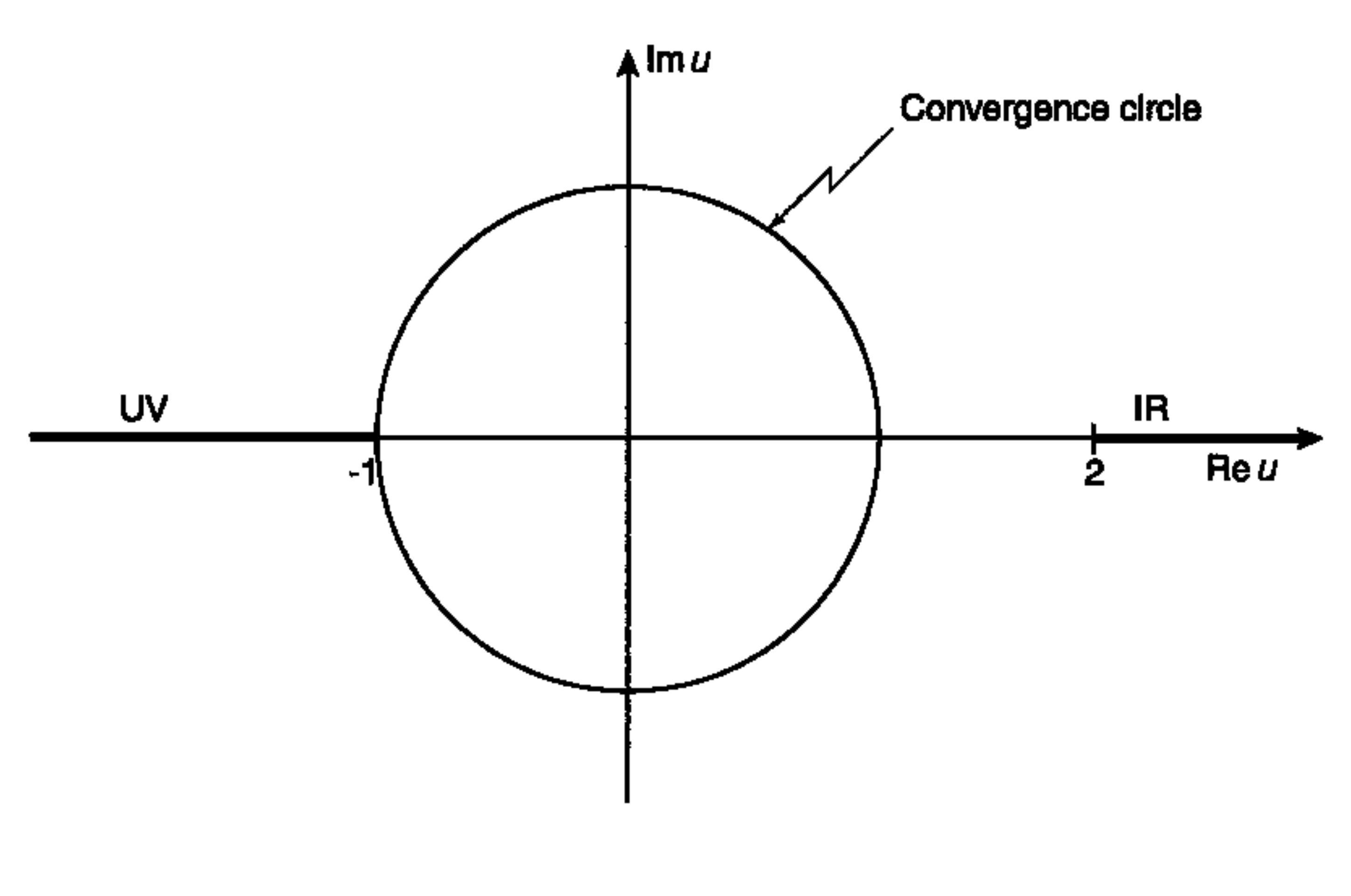}
\caption{Borel plane of the Adler function. The circle indicates the convergence domain of the series (\ref{eq:B}).  
\label{fig:Borel}}
\end{figure}

The large-order increase of the coefficients of the perturbation series is encoded  in the singularities of the Borel transform in the complex $u$ plane.  As shown in Fig. \ref{fig:Borel},  $B_{\wh D}(u)$ has singularities at integer values of $u$ on the semiaxes $u\ge 2$ (IR renormalons and instantons, which we shall neglect in the present analysis since are situated at larger $u$) and $u\le -1$ (UV renormalons).  In the large-$\beta_0$ limit  the singularities are poles, but beyond this limit they are branch points, requiring the introduction of two  cuts along the lines $u\geq 2$ and $u\leq -1$. Apart these cuts,  it is assumed that no other singularities are  present in the complex $u$ plane \cite{Mueller1985}.

\begin{figure}
\includegraphics[width=6cm, height=4.2cm]{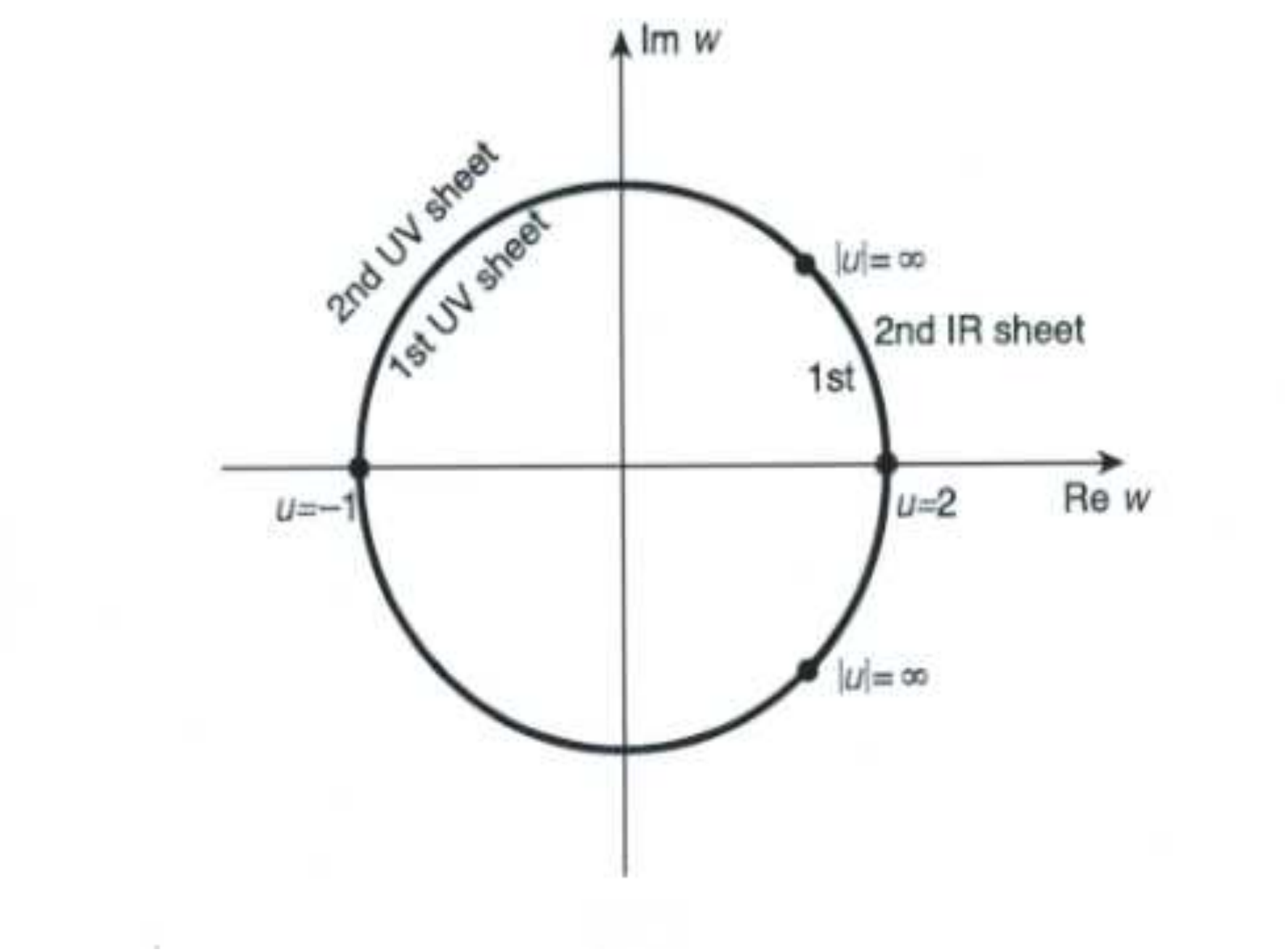}\hspace{0.5cm}
\caption{The  $w$ plane obtained by the conformal mapping (\ref{eq:w}). The IR and UV renormalons are mapped on the boundary of the unit disk.  
\label{fig:confBorel}}
\end{figure}

From the definition (\ref{eq:B}), it follows that the function $\wh D(s)$ defined by (\ref{eq:hatD1}) can be recovered formally from the Borel transform by the Laplace-Borel integral representation 
\be\label{eq:Laplace}
\wh D(s)=\frac{1}{\beta_0} \,\int\limits_0^\infty  
\exp{\left(\frac{-u}{\beta_0 a(-s)}\right)} \,  B_{\wh D}(u)\, d u\,.
\ee
Actually, due to the singularities of $ B_{\wh D}(u)$ for $u\ge 2$, the  integral (\ref{eq:Laplace}) is not defined and requires a regularization. As shown in \cite{Caprini:1999ma}, the Principal Value (PV) prescription, where the integral (\ref{eq:Laplace}) is defined as the semisum of the integrals along two lines, slightly above and below the real positive axis $u\ge 0$, is convenient since it preserves to a large extent the analytic properties of the true function $\wh D(s)$ in the  complex $s$ plane, in particular the absence of cuts on the spacelike axis $s<0$ and Schwarz reflection property. Therefore, we shall adopt this prescription in what 
 follows. 

The singularities of $B_{\wh D}(u)$ set a limitation on the convergence region of the power expansion (\ref{eq:B}): this series converges only inside  the circle $|u|=1$ shown in  Fig. \ref{fig:Borel},  which passes through the first UV renormalon. 
As it is known, the domain of convergence of a power series in the complex plane can be increased by expanding the function in powers of another variable,  which  performs the conformal mapping of the original plane  (or a part of it) onto a disk. In the next section we shall apply this method to the Adler function. 

\section{Nonpower expansions of the Adler function}\label{sec:conf}

The method of conformal mappings  was  introduced  in particle physics in  \cite{CiFi, Frazer, CiCiFi} for improving the convergence of  the expansions of scattering amplitudes in powers of various kinematical variables.  By expanding the amplitude in powers of the function that maps the original analyticity domain onto a unit disk, the new series converges in a larger region,  well beyond the convergence domain of the original expansion, and moreover has an increased  asymptotic convergence rate at points lying inside this domain. The conformal mappings are known actually in mathematics as one of the techniques for ``series acceleration''. 

An important result proved in  \cite{CiFi, CiCiFi}  is that the asymptotic convergence rate is maximal if  the entire holomorphy domain of the expanded function is mapped onto the unit disk.  We recall		
that the large-order convergence rate of a power series at a point in the complex plane is equal to the quotient $r/R$, where
$r$ is the distance of the point from the origin and $R$ the convergence  radius. 
The  proof given in \cite{CiFi} consists in comparing the magnitudes of the 
ratio  $r/R$ 
for a certain point in different complex planes, corresponding to different
conformal mappings. When the whole analyticity domain  of the
function is mapped on a disk, the value of $r/R$ is minimal \cite{CiFi} (a detailed proof is given in \cite{Caprini:2011ya, Caprini:2019osi}).
This defines an ``optimal conformal mapping'', which achieves the best  asymptotic convergence.

In QCD, since the correlators are singular 
at the origin of the coupling plane \cite{tHooft},  the method cannot be used for the standard
perturbative series\footnote{The conformal mapping of the coupling plane was nevertheless used in Refs. \cite{Seznec:1979ev, ZinnJustin:2010ng}, where it was assumed that the singularity is  shifted away from the origin by a certain amount at each finite perturbative order, and tends to the origin only for an infinite number of terms.}.  However, the conditions of applicability are satisfied by the Borel transforms such as $B_{\wh D}(u)$, which are holomorphic in a region containing the origin  $u = 0$  of the Borel complex plane. Thus, the expansion (\ref{eq:B}) in powers of  the Borel variable $u$ can be reexpressed as an expansion in powers of a different variable, which achieves the conformal mapping of the $u$ plane onto the unit disk.

 As shown for the first time in \cite{Caprini:1998wg}, the optimal mapping, which ensures 
the  convergence of the power  series  in the entire doubly-cut Borel plane,  is given by the function 
\begin{equation}\label{eq:w}
\tilde w(u)=\frac{\sqrt{1+u}-\sqrt{1-u/2}}{ \sqrt{1+u}+\sqrt{1-u/2}},
\end{equation}
whose inverse reads
\beq\label{eq:uw}
\tilde u(w)=\frac{8 w}{3-2 w+3 w^2}= \frac{8 w}{ 3 (w-\zeta) (w-\zeta^*)}\,,
 \end{equation}
where $\zeta= (\sqrt{2}+i)/(\sqrt{2}-i)$ and its complex conjugate  $\zeta^*$
are the images of $u=\infty$ on the unit circle in the $w$ plane.

One can check that the function $\tilde w(u)$  maps the complex  $u$ plane cut along the real axis for $u\ge 2$ and $u\le -1$ onto the interior
of the circle $\vert w\vert\, =\, 1$ in the complex plane $w\equiv \tilde w(u)$,  such that  the origin $u=0$ of the $u$ plane
corresponds to the origin $w=0$ of the $w$ plane, and the upper (lower) edges of the cuts are mapped onto the upper
(lower) semicircles in the  $w$ plane (see Fig. \ref{fig:confBorel}). 
By the  mapping (\ref{eq:w}), all  the singularities of the Borel transform, the  UV and IR  renormalons, are pushed on the boundary of the unit disk in the $w$  plane, all at equal distance from the origin. Consider now the expansion of $B_{\widehat D}(u)$ in powers of the variable $w$: 
\be\label{eq:Bw}
B_{\widehat D}(u)=\sum_{n\ge 0} c_n \,w^n, \quad\quad w = \tilde w(u),
\ee
 where the coefficients $c_{n}$ can be obtained from  the coefficients $b_{k}$,
$k\leq n$, using Eqs. (\ref{eq:B}) and  (\ref{eq:w}).   By expanding $B_{\widehat D}(u)$ according to (\ref{eq:Bw}) one makes full use of its
holomorphy domain, because the known part of it
(the first Riemann sheet) is
mapped onto the convergence  disk.  Therefore, the series (\ref{eq:Bw}) converges in the whole $u$ complex plane up to the cuts, i.e., in a much larger domain than the original series (\ref{eq:B}).  Moreover, according to the results mentioned above,  this expansion has the best asymptotic convergence rate compared to other expansions, based on conformal mappings which map a part of the  holomorphy domain onto the unit disk.

By inserting the  expansion (\ref{eq:Bw}) in the Borel-Laplace integral (\ref{eq:Laplace}), we obtain a new perturbative series for the Adler function, of the form
\cite{Caprini:1998wg, Caprini:2000js,Caprini:2001mn}:
\begin{equation}
\wh D(s)= \sum_{n\ge 0} c_{n} {\cal W}_{n}(a(-s)), 
\label{eq:cW}
\end{equation}
where the functions ${\cal W}_n(a)$ are defined as 
\begin{equation}\label{eq:Wn}
{\cal W}_{n}(a)=\frac{1}{\beta_0} {\rm PV}\int\limits_0^\infty\, e^{-u/(\beta_0 a)}\, (\tilde w(u))^n \,du.
\end{equation}

We emphasize that the Principal Value prescription in the definition of the expansion functions preserves to a better extent than other prescriptions the analyticity in the momentum plane and Schwarz reflection property. Preserving analyticity is important in physical applications, which require the analytic continuation of perturbative QCD from 
the spacelike axis to the timelike axis, where measurements are available. 

The expansion can be further improved by exploiting  the fact that the  nature of the leading singularities of $B_{\wh D}(u)$ in the Borel plane is known: near the first branch points $u=-1$ and $u=2$,  $B_{\widehat D}(u)$ behaves like
\begin{equation}\label{eq:gammapowers}
 B_{\widehat D}(u) \sim \frac{r_1}{(1+u)^{\gamma_{1}}} \quad{\rm and} \quad  B_{\widehat D}(u)  \sim \frac{r_2}{(1-u/2)^{\gamma_{2}}}, 
\end{equation}
respectively, where the residues $r_1$ and $r_2$ are not known, but the exponents
$\gamma_1$ and $\gamma_2$ have been calculated \cite{Mueller1985, Mueller:1993pa, Beneke:2008ad}. We shall use the expressions \cite{Beneke:2008ad} 
\beq\label{eq:gamma}
\gamma_1=2-2\,\frac{\beta_2}{\beta_1^2},\quad\quad \gamma_2=1+4\,\frac{\beta_2}{\beta_1^2},
\eeq
involving the first coefficients of the $\beta$ function given in (\ref{eq:betai}). 
For $n_f=3$, when $\beta_1=9/2$ and $\beta_2=8$, (\ref{eq:gamma}) gives
\begin{equation}\label{eq:gamma12}
\gamma_1 = 1.21,    \quad\quad   \gamma_2 = 2.58 \,. 
\end{equation}
 Using (\ref{eq:w}), it is easy to check that 
\bea
(1+u)^{\gamma_1}&\sim & (1+w)^{2\gamma_1}, \quad \text{for}\, u\sim -1\nonumber\\
(1-u/2)^{\gamma_2}&\sim &  (1-w)^{2\gamma_2}, \quad \text{for}\, u\sim 2.
\eea
It follows that the product $B_{\wh D}(u) (1+w)^{2\gamma_{1}} (1-w)^{2\gamma_{2}}$ is finite at $u=-1$ and $u=2$. Actually, the product has still singularities (branch points)  at  $u=-1$ and $u=2$, generated by the terms of $B_{\wh D}(u)$ which are holomorphic at these points, but they are milder than the original ones (the singularities are ``softened'').  It is clear that the optimal variable for the expansion of the product is still the conformal mapping (\ref{eq:w}), which depends only on the position of the first singularities.  Using this remark, we shall adopt the expansion\footnote{The factorisation of the dominant IR renormalon in the Borel plane  was used for the first time for the Adler function in \cite{Soper:1995ns} and for other correlators in \cite{Pineda:2001zq} .} 
\be\label{eq:Bw1}
 B_{\widehat D}(u)=\frac{1}{(1+w)^{2\gamma_{1}} (1-w)^{2\gamma_{2}}} \sum_{n\ge 0} {\widetilde c}_n\, w^n,
\ee
 proposed in \cite{Caprini:2009vf}.  Actually, as emphasized in  \cite{Caprini:2009vf, Caprini:2011ya}, while the optimal conformal mapping (\ref{eq:w}) is unique, the factorization of the singular factors is not. The problem was investigated in detail in \cite{Caprini:2011ya}, where extensive numerical tests indicated the good properties of the expansion (\ref{eq:Bw1}), where the singular factors are simple functions of the variable $w$. In the present paper we  shall adopt the expansion (\ref{eq:Bw1}) and account for other possibilities (for instance, multiplication by the factors $(1+u)^{\gamma_1}(1-u/2)^{\gamma_2}$) in the assessment of the theoretical uncertainty.

By inserting the expansion (\ref{eq:Bw1}) in the Borel-Laplace integral (\ref{eq:Laplace}), we  define a new perturbative series for the Adler function:
\be\label{eq:cWtilde}
\wh D(s)=\sum\limits_{n\ge 0} {\widetilde c}_n {\widetilde W}_n(a(-s)), 
\ee
where the expansion functions are
\be\label{eq:Wntilde}
\hspace{-0.05cm}{\widetilde W}_n(a)=\frac{1}{\beta_0}{\rm PV} \int\limits_0^\infty\!  \frac{{\rm e}^{-\frac{u}{\beta_0 a}}(\tilde w(u))^n}{(1+\tilde w(u))^{2\gamma_{1}} (1-\tilde w(u))^{2\gamma_{2}}}du.
\ee

We note that the expansion functions (\ref{eq:Wn}) and  (\ref{eq:Wntilde}) are no longer powers of the coupling, as in the standard perturbation theory, and exhibit a complicated dependence on $a$. To emphasize this fact, as in \cite{Abbas:2013usa}, we refer to the new expansions  (\ref{eq:cW}) and  (\ref{eq:cWtilde}) as to  ``nonpower expansions''. 

By construction, when reexpanded in powers of $a$, the series (\ref{eq:cW}) and  (\ref{eq:cWtilde}) reproduce the known low-order perturbative coefficients $c_{n,1}$ of the expansion (\ref{eq:hatD1}), given in (\ref{eq:cn1}) and (\ref{eq:c51}).  
On the other hand, as will be argued in the next section, these expansions go beyond standard perturbation theory, allowing to  recapture  nonperturbative features of the expanded function.

\section{Nonperturbative features from perturbation theory}\label{sec:resurg}
\subsection{Properties of the nonpower expansions}\label{sec:prop}
We consider first the analyticity properties of the expansion functions   ${\cal W}_{n}(a)$  in the complex $a$ plane (we recall that $a$ is related to the strong coupling by $a=\alpha_s(-s)/\pi$). It is known that the analytic properties of the QCD correlators in the coupling
constant plane are far from trivial. In \cite{tHooft} it was proved that the multiparticle branch points in the spectral functions at high energies show their presence,  via the  renormalization group equations, in a complicated
accumulation of singularities near the point $a=0$. Since the proof uses a
nonperturbative argument (the existence of multiparticle hadronic states),  it is not possible to see this feature in the standard truncated  perturbation theory: indeed, the expansions in powers of the strong coupling $a$, truncated at finite orders, are holomorphic at the origin  of the  complex $a$ plane and cannot reproduce the singularity of the exact correlator at this point.

For the nonpower expansion functions  ${\cal W}_{n}(a)$ defined in (\ref{eq:Wn}) one expects a more complex structure in the $a$ plane, even after the regularization of the integral by the PV prescription. In \cite{Caprini:2001mn} it was shown that the functions ${\cal W}_{n}(a)$ can be represented in the complex $a$ plane as
 \bea\label{eq:Wnaul}{\cal W}_{n}(a)=\int\limits_0^\infty\, e^{-t}\,  [\tilde w( t a)]^n\, d t \mp i\,e^{-\frac{2}{a}}\, \int\limits_0^\infty\, 
e^{-t}\,  f_n( t a)\, d t\,, \nonumber\\ \text{Im}\, a \gtrless 0,\quad\quad\eea 
where the functions $f_n(a)$ are defined in Eq. (24) of Ref. \cite{Caprini:2001mn}. As further proved in \cite{Caprini:2001mn}, the representation  (\ref{eq:Wnaul}) implies that the functions ${\cal W}_{n}(a)$  are 
analytic functions of real type, i.e., they satisfy the Schwarz reflection property  ${\cal W}_{n}(a^*)=  ({\cal W}_{n}(a))^*$, in the whole complex $a$ plane, except for a cut along the real negative axis $a<0$ and an essential singularity at $a=0$. 
Therefore, the expansion (\ref{eq:cW}), even if truncated at a finite order,  exhibit a feature of the full correlator, namely its singularity at the origin of the $a$ plane. 
  
It is useful to note that the new expansions, when reexpanded in powers of $a$, contain an infinite number of terms, even if the expansions themselves are truncated at finite orders.  Thus, the truncated expansions  (\ref{eq:cW}) and  (\ref{eq:cWtilde}) go beyond standard finite-order perturbation theory.  This remark will be useful below. 

We can actually investigate in more detail the perturbative expansion of the functions ${\cal W}_n(a)$ themselves. Since these functipns  have singularities at $a=0$,  their Taylor expansions around the origin will be  divergent series.  We take first $a$
real and positive, when the  functions ${\cal W}_n(a)$ are well defined and have bounded magnitudes.  By applying  Watson's
lemma \cite{Watson} (see also \cite{Jeff} and \cite{Caprini:2009nr}), it was shown in \cite{Caprini:2001mn} that ${\cal W}_{n}(a)$ can be expressed as
\beq\label{eq:Wna}
{\cal W}_{n}(a) = \sum\limits_{k=n}^N \xi_k^{(n)} k!\, a^k + M_n (N+1)!\, a^{N+1} + O\left({\rm e}^{-\frac{X}{ a}}\right), 
\eeq 
where $N$ is a positive integer, $M_n$ is independent of $N$, $X$ is an arbitrary positive parameter less than 1 and $\xi_k^{(n)}$ are defined by the Taylor expansions 
\begin{equation}\label{eq:wnser}
(\tilde w(u))^n=\sum\limits_{k=n}^\infty \xi_k^{(n)} u^k,\quad\quad n\ge 1.
\end{equation}
The expression (\ref{eq:Wna}) implies that 
\beq\label{eq:Wnasim}
R_N^{(n)}\equiv{\cal W}_{n}(a) - \sum\limits_{k=n}^N \xi_k^{(n)} k!\, a^k=o(a^{N}),\quad  a\to 0_+
\eeq
which is the definition of an asymptotic expansion \cite{Jeff},  so we can write using a standard notation
\begin{equation}
{\cal W}_{n}(a) \sim \sum\limits_{k=n}^\infty \xi_k^{(n)} k! a^k,\quad\quad a\to 0_+.\nonumber
\end{equation} 
We recall that, while the convergence of a series can be established or disproved only from the knowledge of its coefficients, for an asymptotic expansion one needs to know both the function and the coefficients. On the other hand, while a convergent series has a unique sum, the coefficients of an asymptotic series do not determine  the function uniquely. More information, like for instance analyticity in a region of the complex plane near $a=0$, is necessary in general to ensure uniqueness.

As shown in \cite{Caprini:2001mn}, the representation (\ref{eq:Wna}) is independent of the prescription adopted for the Borel-Laplace integral. We note that the first term of each ${\cal W}_{n}(a)$ is
proportional to $n!\, a^n$ with a positive coefficient,  thereby  retaining a
fundamental property of perturbation theory. But the series (\ref{eq:Wna}) 
are divergent: indeed, since the expansions (\ref{eq:wnser}) have the convergence
radii equal to 1, there are for any $R > 1$ infinitely many $k$ such that  
$|\xi^{(n)}_k|>R^{-k}$ \cite{Jeff}. Actually, the divergence of the series 
(\ref{eq:Wna}) is not surprising, in view of the singularities of the functions  ${\cal W}_{n}(a)$ at the origin of the $a$ plane. 

 For illustration, we
give below the expansions of the first functions ${\cal W}_n(a)$ defined in (\ref{eq:Wn}) for $n\ge 1$ (note that ${\cal W}_{0}(a)=a$): 
\begin{widetext}
 \bea \label{eq:Wnseries}
{\cal W}_{1}(a) &\sim&0.844\, a^2 - 0.949\, a^3 + 5.206\, a^4 - 27.932 \,a^5 + 249.61\,a^6 - 
 2535.85\, a^7 + 32810.9 \,a^8 - 485719 \, a^9 +\ldots \nonumber \\ 
{\cal W}_{2}(a)&\sim& 1.424\, a^3 - 4.805\, a^4 + 40.546\, a^5 - 334.502\, a^6 + 3864.71\, a^7 - 
 50084.5\, a^8 + 777892\, a^9 +\ldots \nonumber \\
{\cal W}_3(a) &\sim &3.604 \,a^4 - 24.327\, a^5 + 290.789\, a^6 - 3367.59\, a^7 + 49042.7\, a^8 - 
 785848\, a^9 +\ldots   
\eea \end{widetext} 
The higher powers of $a$  become quickly more and more important in (\ref{eq:Wnseries}), for instance, the 12th-order coefficients are $3.9\times 10^{12}$, $-5.9\times 10^{12}$ and $ 6.6\times 10^{12}$, respectively, the coefficients  exhibiting a
factorial growth.

 The expansion functions $\widetilde {\cal W}_n(a)$ defined in (\ref{eq:Wntilde}) have similar properties: they are singular at the origin of the $a$ plane and their  expansions in powers of $a$ are divergent (in particular, the coefficient of the first term is identical to that of (\ref{eq:Wna}), i.e.,  $\xi_n^{(n)} n!$). The expansions of the first functions $\widetilde {\cal W}_n(a)$ have the form: 
\begin{widetext}
 \bea \label{eq:Wntseries}
\widetilde {\cal W}_{0}(a)&\sim&   a + 2.312\, a^2 + 8.140\, a^3 + 31.088\, a^4 + 213.55\, a^5 + 
 980.805\, a^6 + 13677.8\, a^7 + 30900.7\, a^8 + 1.95\times 10^6\, a^9                  +\ldots \nonumber \\ 
\widetilde {\cal W}_{1}(a)&\sim&0.844\, a^2 + 2.952\, a^3 + 19.227\, a^4 + 78.770\, a^5 + 956.331\, a^6 + 
 2677.55\, a^7 + 104194\, a^8 - 308869\, a^9 +\ldots \nonumber \\ 
\widetilde {\cal W}_{2}(a)&\sim& 1.424\, a^3 + 5.069\, a^4 + 65.649\, a^5 + 185.647\, a^6 + 5748.59\, a^7 - 
 7196.39\, a^8 + 1.04 \times 10^6\, a^9 +\ldots \nonumber \\ 
\widetilde {\cal W}_{3}(a)&\sim& 3.604\, a^4 + 9.001\, a^5 + 302.958\, a^6 - 63.591\, a^7 + 44720.9\, a^8 - 
 305830\, a^9 +\ldots  
\eea \end{widetext} 

We conclude that, unlike the expansion functions $a^n$ of the standard perturbation theory, which are holomorphic at $a=0$,  the nonpower expansion functions ${\cal W}_n(a)$ and $\widetilde {\cal W}_n(a)$ are singular at $a=0$ and admit divergent expansions in powers of $a$, resembling from this point of view the expanded function $ \wh D$ itself. 

On the other hand, as proved in \cite{Caprini:2000js}, the new expansions  (\ref{eq:cW}) and  (\ref{eq:cWtilde}) have a tamed behavior at high orders and, under certain conditions, they  may even converge in a domain of the $s$ plane. Crucial for the proof is the large-order behavior of the functions ${\cal W}_{n}(a)$ at large $n$,  investigated  in  
\cite{Caprini:2000js, Caprini:2001mn} by the technique of saddle points. Omitting the details given in \cite{Caprini:2000js}, we quote  the
asymptotic  behavior of ${\cal W}_n(a)$ for $n\to \infty$:
\bea\label{eq:Wnpsaddle} {\cal W}_n(a)&\approx& n^{\frac{1}{ 4}}{e}^{-2^{3/4} (n/a)^{1/2}}\\
&\times& [\zeta^n 
{e}^{-2^{3/4} i \,(n/a)^{1/2}} +  (\zeta^*)^n
 {e}^{2^{3/4} i\, (n/a)^{1/2}}], \nonumber \eea where $\zeta$ was
defined below (\ref{eq:uw}). 
The estimate  (\ref{eq:Wnpsaddle})  is valid in the complex $a$ plane, for  $a=|a|
{\rm e}^{{\rm i}\psi}$ with $\psi$ restricted by
\begin{equation}
|\psi|<\pi/6.
\label{eq:psi}
\end{equation}

The convergence of the expansion (\ref{eq:cW}) depends on the ratio
\begin{equation}\label{eq:ratio}
\bigg\vert\frac{c_n {\cal W}_n(a)}{ c_{n-1} {\cal W}_{n-1}(a)}\bigg\vert\,.
\end{equation}
As shown in \cite{Caprini:2000js, Caprini:2001mn}, if the coefficients $c_n$ satisfy the condition
 \begin{equation}\label{eq:cnb} |c_n| < C {\rm e}^{\epsilon
n^{1/2}}\, \end{equation}
with $C>0$ for any $\epsilon >0$, the expansion (\ref{eq:cW}) converges for $a$ complex in the domain
\begin{equation}\label{eq:domain}
{\rm Re}[(1\pm i) a^{-1/2}]>0,
\end{equation}
which is equivalent to $ |\psi|\leq \pi/2-\delta$, for any $\delta>0$.
Since the condition   (\ref{eq:psi}) is more restrictive, it follows that, if the 
condition (\ref{eq:cnb}) is satisfied,  
the series (\ref{eq:cW}) converges in the sector defined by (\ref{eq:psi}).

\begin{table}[t]
\caption{Adler function $\wh D(-m_\tau^2)$ predicted by the ``reference model'' (see text) for $\alpha_s(m_\tau^2)=0.32$, calculated with the perturbative standard expansion (\ref{eq:hatD1}) and the nonpower expansions (\ref{eq:cW}) and (\ref{eq:cWtilde}), for various truncation orders $N$. Exact value: $\wh D(-m_\tau^2)=0.137706$.} \vspace{0.15cm}
\label{tab:1}
 \renewcommand{\tabcolsep}{0.55pc} 
\renewcommand{\arraystretch}{1.15} 
\begin{tabular}{|l|c|c| c | }\hline\hline
$N$ \,\,\,    &\,\,\,Eq. (\ref{eq:hatD1})  &\,\,Eq. (\ref{eq:cW})& Eq. (\ref{eq:cWtilde}) \\\hline
10 & 0.155429 & 0.142247& 0.137763       \\   
 11 & 0.149068  &0.139757 & 0.137733     \\
 12 & 0.191213 & 0.137235& 0.137700   \\
 13 & 0.114491 & 0.135647& 0.137712     \\
 14 & 0.417809& 0.135401& 0.137729    \\
 15 & -0.442007& 0.136258& 0.137724   \\
 16 & 2.80676& 0.137549& 0.137715     \\
 17 & -8.76330 & 0.138553& 0.137714   \\
 18 & 37.9988 & 0.138851& 0.137716     \\
 19 & -154.7999& 0.138470& 0.137714     \\
 20 & 700.409 & 0.137788& 0.137711   \\
 21 & -3248.105 & 0.137259& 0.137709     \\
 22 & 15993.08 & 0.137139& 0.137709    \\
 23 & -81886.8 & 0.137384 &0.137709     \\
 24 & 439277.8 & 0.137744& 0.137707     \\
 25  &-2.45 $\times 10^6$&0.137973 &0.137706     \\[0.04cm]
\hline\hline \end{tabular}\end{table}

The validity of the condition (\ref{eq:cnb}) in QCD cannot be proved formally. Instead, the convergence of the series based on the conformal mapping of the Borel plane was confirmed numerically in realistic models of the Adler function inspired from real QCD. These models, proposed for the first time in   \cite{Beneke:2008ad}, parametrize the Borel transform $B_{\wh D}(u)$ as a sum of IR and UV renormalon contributions and a regular part, which satisfy renormalization-group invariance and reproduce the known low-order coefficients of the expansion (\ref{eq:hatD1}). 
As shown in \cite{Caprini:2009vf, Caprini:2011ya,   Caprini:2018agy}, the improved expansions provide a much better approximation that the standard PT, up to high orders.

For illustration, we consider here the perturbative calculation of the Adler function on the spacelike axis, using  the ``reference model'' proposed in  \cite{Beneke:2008ad} and an alternative model,  proposed in \cite{Caprini:2011ya}, with a smaller residue of the first IR renormalon (these models are summarized in Appendix A of  \cite{Caprini:2019kwp}). The exact value of the Adler function is obtained by inserting the Borel transform described by each model\footnote{Note that in the conventions used in this paper the expression of $B_{\wh D}(u)$ given in  Appendix A of \cite{Caprini:2019kwp} must be  multiplied by $\pi$.} in the PV-regulated Borel-Laplace integral (\ref{eq:Laplace}).  On the other hand, from the perturbative
coefficients of these models, calculated exactly to any order, one can obtain the standard perturbation expansion  (\ref{eq:hatD1}) and construct also the improved ones, given in Eqs. (\ref{eq:cW}) and (\ref{eq:cWtilde}).

\begin{table}[t]
\caption{The same as in Table \ref{tab:1} for the alternative model (see text). Exact value: $\wh D(-m_\tau^2)= 0.139136.$ } \vspace{0.15cm}
\label{tab:2}
 \renewcommand{\tabcolsep}{0.55pc} 
\renewcommand{\arraystretch}{1.15} 
\begin{tabular}{|l|c| c |c| }\hline\hline
$N$ \,\,\,    & Eq. (\ref{eq:hatD1})  & Eq. (\ref{eq:cW}) & Eq. (\ref{eq:cWtilde}) \\\hline
  10 & 0.146532& 0.140557 & 0.140773  \\
  11& 0.135890& 0.140106&  0.141022 \\
  12& 0.171622& 0.139288& 0.140710  \\
  13& 0.084619&0.138610 &0.139955  \\
  14& 0.370536& 0.138501 &0.139723  \\
  15& -0.521082&  0.138830  &0.139711  \\
  16& 2.667712&0.139219  &0.139499  \\
  17& -9.02819&0.139463  &0.139326  \\
  18& 37.47991& 0.139534  & 0.139312  \\
  19& -155.9579& 0.139429  &0.139297  \\
  20& 698.025& 0.139205 &0.139239 \\
  21& -3254.873& 0.139011  &0.139206  \\
  22& 15981.426&0.138967 &0.139205  \\
  23& -81944.79& 0.139050&0.139197 \\
  24& 439264.5&0.139159 &0.139176  \\
  25& -2.45 $\times 10^6$&  0.139225 & 0.139165
 \\[0.04cm]
\hline\hline \end{tabular}\end{table}

In Table \ref{tab:1} we present the predictions of these expansions truncated at the finite order  $N$ for the Adler function $\wh D(s)$ given by the ``reference model'' mentioned above. We take the point $s=-m_\tau^2$ on the spacelike axis, far from the hadronic thresholds, where perturbative QCD can be applied. The calculations have been done with  $\alpha_s(m_\tau^2)=0.32$. Since we are interested in the high-order behavior of the expansions, 
we show the results for $N$ larger than 10.

 One can see that, while the standard expansion  (\ref{eq:hatD1}) wildly diverges, the improved expansions  converge to the exact value $\wh D(-m_\tau^2)= 0.137706$ predicted by the model.  This pattern is preserved to higher orders: for 
instance, for $N=40$, the standard expansion  predicts $2.39 \times 10^{19}$, while the improved expansions give 0.137727 and 0.137706, respectively (the expansion (\ref{eq:cWtilde}) reproduces actually the exact value to 7 digits). Moreover, the results show that the  explicit factorization of the first singularities of the Borel transform  improves the approximation both at low and high orders. 

In Table \ref{tab:2} we present similar results for the alternative model given in Appendix A of  \cite{Caprini:2019kwp}, for which the exact value is $\wh D(-m_\tau^2)=0.139136$. The good convergence pattern of the nonpower expansions is illustrated in the last two columns, in contrast to the divergence of the standard PT shown in the first column.  The same features are preserved at higher orders: for instance, for $N=40$ the results are $2.39\times 10^{19}$, 0.139142 and 0.139019, respectively.

The good convergence properties of the new expansions, discussed above, will play an important role in the interpretation of these expansions as an alternative to the standard OPE for recapturing nonperturbative features of the expanded function.

\subsection{Nonpower expansions versus standard OPE}\label{sec:compar}
From the above discussion, it follows that the expansions (\ref{eq:cW}) and (\ref{eq:cWtilde}) exhibit crucial nonperturbative features of the expanded function and allow to recover this function from its perturbative coefficients. 
It is of interest to look in parallel at the properties of the standard OPE, which, as mentioned already, is an example of the transseries concept applied to QCD. 

 We recall that in the frame of OPE, largely used in QCD phenomenology since its proposal in Ref. \cite{Shifman:1978bx}, the representation of the Adler function 
\be\label{eq:OPE}
\widehat{D}(s) \sim \sum\limits_{n= 1}^N c_{n,1}\, [a(-s)]^n + \sum\limits_{k=1}^K \frac{d_k}{(-s)^k}
\ee
contains, besides the truncated perturbative expansion,  a series of ``power corrections'', with coefficients $d_k$ involving both perturbative factors depending logarithmically on $s$ and nonperturbative condensates. 

Despite its great popularity, one must keep in mind that OPE expansion, when generalized 
to include  power corrections, is  an assumption. As it is known,  the validity of 
the OPE is  only proven rigorously within  perturbation theory, and is postulated in the 
nonperturbative framework.  This fact is emphasized in many places (see for instance \cite{Bali:2014sja, Ayala:2019uaw}).

For the present discussion, the crucial remark is that $1/(-s)^k$ can be written approximately  as  $\exp [-k/(\beta_0 a(-s))]$, where $a(-s)$ is the expansion parameter of the first series in (\ref{eq:OPE}),  calculated by solving the renormalization group equation (\ref{eq:rge}). Therefore, the power corrections in the OPE can be identified with the  nonanalytic terms, exponentially small in the expansion parameter of a divergent series, which  must be added to it in order to recover the expanded function.   On the other hand, as discussed below (\ref{eq:Wnaul}), the expansion functions ${\cal W}_n(a)$ (and $\wt {\cal W}_n(a)$, actually)  exhibit too  singularities near the origin of the complex $a$ plane. Thus, both OPE and the nonpower expansions based on the conformal mapping of the Borel plane incorporate a nonperturbative feature of the exact Adler function, although neither can reproduce exactly the complicated singularity structure of this function near $a=0$, found in \cite{tHooft}.

The Borel plane provides another argument for the similarity of the two approaches. As discussed in the mathematical literature \cite{BerryHowls, Howls}, in the so-called hyperasymptotic approach the transseries account for the singularities in the Borel plane, which the ordinary asymptotic expansion fails  to deal with.  Indeed, the action of
taking the Laplace-Borel transform (\ref{eq:Laplace}) over an infinite range,  beyond the finite
radius of convergence of the expansion (\ref{eq:B}) of the Borel transform, generates the
divergent asymptotic expansion of the Adler function [the first series in (\ref{eq:OPE})]. In order to overcome this, the hyperasymptotics approach includes additional terms (the second series in (\ref{eq:OPE})), which is equivalent in a certain sense to an analytic continuation of the Borel transform to the neighbourhood of
the distant singularities \cite{Howls}. This  allows the function to ``resurge'', or to be asymptotically remodelled.
On the other hand, the expansions  (\ref{eq:Bw}) and  (\ref{eq:Bw1}), based on conformal mapping of the Borel plane,  converge in the whole $u$ plane up to the cuts, achieving in a manifest way the analytic continuation outside the circle of convergence of the series (\ref{eq:B}). 

Therefore, the transseries approach and the method of conformal mapping represent alternative ways to effectively perform the analytic continuation in the Borel plane, in order to recover the expanded function when its asymptotic perturbative expansion diverges. 
 This can be seen from the fact that nonperturbative features similar to those introduced explicitly  in the standard OPE are contained in an implicit way in the  expansions (\ref{eq:cW}) or (\ref{eq:cWtilde}).   We can make therefore the conjecture that the nonpower  expansions provide by themselves a consistent way of recapturing the exact function, without the need of additional power corrections, being an alternative to the standard OPE.  Below we present two additional arguments in favour of this conjecture. 

First, we emphasize that we do not assume that the nonperturbative condensates are zero. We recall however that these terms have been defined within OPE. Moreover, as discussed in recent analyses \cite{Bali:2014sja, Ayala:2019uaw, Ayala:2019hkn}, the nonperturbative terms depend on the perturbative part, in particular on the truncation order of the perturbation series. But the  new expansions defined here,  even if truncated at finite orders,
contain an infinite number of terms when reexpanded in powers of $a$. So, if one may think to supplement them by additional power corrections, their interpretation in terms of condensates will be hard to give. 

More importantly, the new expansions, which are obtained by a systematic  
mathematical method, are shown to converge under some conditions (whose validity is 
expected to hold in QCD), and the convergence is checked numerically  on models inspired from QCD. 
There are  no  reasons for adding to
a convergent series arbitrary terms (in the transseries approach, such terms are 
necessary for recapturing a function from its divergent asymptotic expansion). 
In conclusion, there are no mathematical arguments  for supplementing the nonpower expansions by 
additional, arbitrary power corrections.

The conjecture formulated above implies in particular that the difference between the predictions of the nonpower expansions and the pure perturbative part of OPE (the first series in (\ref{eq:OPE}) should be of the order of magnitude of the power corrections.  Below we make a rough numerical test of this expectation, using for illustration the Adler function  $\wh D (s)$ at the spacelike point $s=-m_\tau^2$.

By inserting the known coefficients $c_{n,1}$ given in (\ref{eq:cn1}) and the central estimate of $c_{5,1}$ from  (\ref{eq:c51}) in the standard PT expansion (\ref{eq:hatD1}), we obtain for $\alpha_s(m_\tau^2)=0.32$ the value $\wh D_\text{PT}(-m_\tau^2)=0.1339$. 
On the other hand, for the same input the first five coefficients $\wt c_n$ in (\ref{eq:Bw1}) are:
\bea\label{eq:cntilde}
&&\wt c_0=1, \quad \,\wt c_1=- 0.80,\quad \wt c_2= 0.41,\\
&&\wt c_3= 8.66,\quad\wt c_4= 1.75\pm 4.19, \nonumber
\eea
where the uncertainty of $\wt c_4$ is due to the uncertainty of $c_{5,1}$ quoted in (\ref{eq:c51}. Then the expansion (\ref{eq:cWtilde}) predicts for the same coupling  the central value  $\wh D(-m_\tau^2)=0.1384$, larger by 0.0045 than the standard PT value. 

For the power corrections, the analyses made in \cite{Braaten:1991qm, Beneke:2008ad} show that the dominant contribution is given by the gluon condensate. Using as in \cite{Beneke:2008ad} the standard historical value  $\langle a G^2\rangle = 0.012\ \text{GeV}^4$, we estimate the  contribution of the the power corrections to $\wh D(-m_\tau^2)$ as $0.006\pm 0.006$, where a conservative error of 100\% was added.  This interval is consistent with the difference of about 0.005 between the standard and the nonpower perturbative expansions, which roughly confirms the conjecture made above. We note  that a more reasonable 
comparison in the spirit of the work done in Refs. \cite{Ayala:2019uaw, Ayala:2019hkn}  would require the truncation 
of the standard perturbative series at the minimal term, when the definition of the nonperturbaive terms 
 is more rigorous. However, 
for the Adler function the minimal term is expected to occur at a 
higher order $n$, which is not yet reached by Feynman-graph calculations.  Therefore, 
a more rigorous estimate is not possible at the present status of knowledge of the Adler function. 

In Sec. \ref{sec:alphas} we shall exploit the consequences of the formalism proposed in this paper for the evaluation of the strong coupling constant $\alpha_s$ from hadronic $\tau$  decay. Before this, we shall investigate in more detail the method of conformal mapping for observables represented by integrals of the Adler function along a contour in the complex $s$ plane.
 
\section{Moments of the spectral function}\label{sec:moments}
 The moments of the spectral function $\text{Im} \Pi(s)$ are defined as weighted integrals of this quantity along the physical region $4 m_\pi^2\leq s\leq m_\tau^2$ of the hadronic decays of the $\tau$ lepton. They are accessed through experiment and play an important role in the extraction of the QCD parameters, in particular the strong coupling $\alpha_s$, from hadronic $\tau$ decays. More generally, the moments are 
defined as \cite{Beneke:2012vb}
\beq\label{eq:delwi}
M_{w_i}(s_0)= 
\frac{2}{\pi}\int\limits_0^{s_0} w_i(s/s_0) \,{\rm Im} \Pi(s+i\epsilon)\,ds,
\eeq 
where $0<s_0\leq m_\tau^2$ and  $w_i(x)$ are arbitrary nonnegative weights. 
We are interested in the  pure perturbative contribution 
to $M_{w_i}$, denoted as $\delta^{(0)}_{w_i}$, obtained by subtracting 
from (\ref{eq:delwi}) the tree values $\delta^{\rm tree}_{w_i}(s_0)$.

If the weights $w_i(s)$ are holomorphic functions in the disk $|s|\leq s_0$, taking into account the analytic properties of $\Pi(s)$ and applying Cauchy theorem one can write  
equivalently (\ref{eq:delwi})  as an integral along a contour in the complex
$s$ plane, chosen for convenience to be the circle $|s|=s_0$. After an integration by parts, the perturbative contribution
 $\delta^{(0)}_{w_i}$ can be written as
\begin{equation}\label{eq:del0}
\delta^{(0)}_{w_i}(s_0)= \frac{1}{2\pi i} \!\!\oint\limits_{|s|=s_0}\!\! \frac{d s}{s} 
W_i(s/s_0) \widehat{D}(s),
\end{equation}
where the weights $W_i(x)$ are defined as 
\beq\label{eq:Wi}
W_i(x)=2\int_x^1 dz \, w_i(z),
\eeq  and $\widehat{D}$ is  the reduced Adler function (\ref{eq:D}). 

  Perturbative QCD is not directly applicable for the evaluation of the observables (\ref{eq:delwi}), since it cannot describe the hadronic thresholds  in the spectral function on the timelike axis. However, the equivalent expression (\ref{eq:del0}) involves the values of the Adler function in the complex plane, where perturbation theory makes sense (especially if the region near the timelike axis is suppressed by a suitable choice of the weight $W_i$).  We can insert therefore in 
(\ref{eq:del0}) the perturbative expansions, either (\ref{eq:hatD}) or (\ref{eq:hatD1}), of the Adler function.

The first alternative is known as fixed-order (FO)  perturbation theory and leads to an expansion of the form
\beq\label{eq:deltaFO}
\delta^{(0)}_{w_i,\text{FO}}(s_0)= \sum\limits_{n\ge 1} d_n\, [a(s_0)]^n
\eeq
where the coefficients $d_n$ are obtained by integrating the $s$-dependent coefficients of  (\ref{eq:hatD}) along the circle, and $a(s_0)$ is the coupling at the scale $\mu^2=s_0$. In the second alternative, known as contour-improved (CI) perturbation theory, the expansion of $\delta^{(0)}_{w_i}(s_0)$ reads
\beq\label{eq:deltaCI}
\delta^{(0)}_{w_i, \text{CI}}(s_0)= \sum\limits_{n\ge 1} c_{n,1}  \frac{1}{2\pi i} \!\!\oint\limits_{|s|=s_0}\!\! \frac{d s}{s} 
W_i(s/s_0) [a(-s)]^n,
\eeq
where the running coupling $a(-s)$ is computed by integrating the equation (\ref{eq:rge}) iteratively along the circle, starting from a given $a(s_0)$.

The comparison between the standard FO and CI perturbative QCD expansions  of the moments has been investigated in \cite{Beneke:2012vb, Boito:2020hvu}, where substantial differences between the two ways of renormalization-group summation have been noticed. Here we are interested in the method of conformal mapping of the Borel plane, which can be applied to improve the expansions of the quantities $\delta^{(0)}_{w_i}$ much like that of the Adler function itself. This problem has been investigated in \cite{Caprini:2009vf, Caprini:2011ya, Abbas:2013usa}. 

For the CI version of summation, the application of the conformal mapping is straightforward: one has simply to insert in  (\ref{eq:del0}) the improved expansions (\ref{eq:cW}) or (\ref{eq:cWtilde}) of  $\widehat{D}(s)$. For the FO version,  one must follow the steps applied in Sec. \ref{sec:Adler} to the Adler function, using now as starting point the expansion (\ref{eq:deltaFO}). We define first the Borel transform
\be\label{eq:Bdelta}
 B_{\delta_{w_i}}(u)= \sum_{n=0}^\infty  b_n'\, u^n,
\ee
where $b_n'$ are related to the coefficients $d_{n}$ by 
\be\label{eq:bndelta}
 b_n'= \frac{d_{n+1}}{\beta_0^n \,n!}.
\ee
Then $\delta_{w_i, \text{FO}}^{(0)}$ is recovered from its Borel transform by the Laplace-Borel integral
\be\label{eq:Laplacedelta0}
\delta_{w_i, \text{FO}}^{(0)}=\frac{1}{\beta_0} \text{PV} \,\int\limits_0^\infty  
\exp{\left(\frac{-u}{\beta_0 a(s_0)}\right)} \,  B_{\delta_{w_i}}(u)\, d u,
\ee
where we adopted the Principal Value anticipating the presence of singularities of the Borel transform $B_{\delta_{w_i}}(u)$ on the integration axis. 

 The  analytic properties of the  Borel transform $B_{\delta_{w_i}}(u)$ defined in (\ref{eq:Bdelta}) in the complex $u$ plane  have been investigated some time ago in \cite{Brown:1992pk} and more recently in \cite{Beneke:2008ad, Caprini:2019kwp,  Boito:2020hvu}.
Inserting  the Laplace-Borel representation (\ref{eq:Laplace}) into the integral (\ref{eq:del0}) and permutting the integrals we obtain
\beq\label{eq:integralphi}
\delta_{w_i, \text{FO}}^{(0)} = \frac{1}{\beta_0} \,\int\limits_0^\infty  d u\,  \,  B_{\wh D}(u)\, \frac{1}{2\pi}\int\limits_{0}^{2\pi} \,  d\phi\,W_i(s/s_0)\,
e^{\frac{-u}{\beta_0 a(-s)}},
\eeq
where $s=s_0 \exp(i(\phi-\pi))$. 

 The integral upon $\phi$ can be performed exactly in the one-loop (large-$\beta_0$) approximation, when (\ref{eq:rge}) implies
\beq\label{eq:oneloop}
\frac{1}{\beta_0 a(-s)}=\frac{1}{\beta_0 a(s_0)} + \ln\left(\frac{-s}{s_0}\right), \nonumber
\eeq
the last term being equal to $i(\phi-\pi)$. Then, the comparison of (\ref{eq:integralphi}) with (\ref{eq:Laplacedelta0}) leads to 
\beq\label{eq:BdeltaiBD}
B_{\delta_{w_i}}(u) = \left[\frac{1}{2\pi} \int_0^{2\pi} d\phi \,W_i(e^{i\phi})\, e^{-iu(\phi-\pi)}\right] B_{\wh D}(u).
\eeq
 The integral can be calculated exactly for polynomial weights. In particular, for $w_i(x)=x^n$, one has \cite{Boito:2020hvu}
\beq\label{eq:BxnBD}
B_{\delta_{x^n}}(u)= \frac{2}{1+n-u}\,\frac{\sin\pi u}{\pi u}\,B_{\wh D}(u).
\eeq
From  this relation it follows that the singularities of $B_{\wh D}(u)$  at integer values of $u$  are partly compensated by the zeros of $\sin\pi u$, except for $u=n+1$.  Thus, for a fixed $n$, $B_{\delta_{x^n}}(u)$ inherits from $B_{\wh D}(u)$ the branch point at $u=n+1$, while the other branch points are weakened by simple zeros. The argument can be extended in a straightforward way to more general polynomial weights.

The relation (\ref{eq:BdeltaiBD}) is valid in the one-loop (or large-$\beta_0$)  approximation for the coupling. As proved in \cite{Boito:2020hvu}, the connection between the Borel transforms remains the same also in the exact case of the full renormalization-group equation (\ref{eq:rge}) in a special RS, known as $C$-scheme, defined in \cite{Boito:2016pwf} and investigated further in \cite{Caprini:2018agy, Caprini:2019kwp, Boito:2020hvu}.  For other RS's, in particular  $\overline{{\rm MS}}$, a relation of the type (\ref{eq:BxnBD}) cannot be proved. As discussed in \cite{Caprini:2019kwp}, the exact nature of the first singularities of the  moments cannot be established exactly, although the large-$\beta_0$  approximation may offer a hint. Therefore, if one wants to write for the moments improved expansions of the form (\ref{eq:cWtilde}), with expansion functions (\ref{eq:Wntilde}),  a conjecture  about the nature of the first singularities  is necessary. In the next section we shall discuss this problem in more detail for a particular moment  of physical interest.

\vspace{-0.2cm}

\section{$\tau$ hadronic width}\label{sec:delta0}
 The ratio $R_\tau$ of the total $\tau$ hadronic branching fraction to the electron branching fraction is expressed in the SM as \cite{Beneke:2008ad}
\be\label{eq:Rtau}
R_\tau= 3 \,S_{\rm EW} (|V_{ud} |^2 + |V_{us}|^2 ) (1 + \delta^{(0)} + \ldots),
\ee
where $S_{\rm EW}$ is an electroweak factor, $V_{ud}$ and $V_{us}$ are CKM  matrix elements, and $\delta^{(0)}$ is a perturbative QCD correction.  As shown in   \cite{Braaten:1988hc, Braaten:1991qm, LeDiberder:1992zhd}, this quantity can be written as a weighted integral of the Adler function along a contour in the complex $s$ plane, taken for convenience to be the circle $|s|=m_\tau^2$. In our normalization, this relation is \cite{Beneke:2008ad}: 
\be\label{eq:delta0}
\delta^{(0)} =  \frac{1}{2\pi i} \oint\limits_{|s|=m_\tau^2}\, \frac{d s}{s}\, W_\tau(s)\,\wh D(s),
\ee
where
\beq\label{eq:Wtau}
 W_\tau(s)=\left(1-\frac{s}{m_\tau^2}\right)^3\,\left(1+\frac{s}{m_\tau^2}\right).
\eeq

Perturbative expansions of $\delta^{(0)}$ improved by the optimal conformal mapping of the Borel plane have been proposed and investigated in 
\cite{Caprini:2009vf, Caprini:2011ya, Abbas:2013usa}, in both CI and FO renormalization-group resummations. 
The improved $\delta_{\rm CI}^{(0)}$ expansion is obtained in a straightforward from (\ref{eq:delta0}) and (\ref{eq:cWtilde}) as
\beq\label{eq:del0CI}
\delta_{\rm CI}^{(0)} =  \frac{1}{2\pi i} \sum_{n\ge 0} {\widetilde c}_n \!\! \oint\limits_{|s|=m_\tau^2}\, \frac{d s}{s} \,W_\tau(s) \, {\widetilde {\cal W}}_n(a(-s)), 
\eeq
with $ {\widetilde {\cal W}}_n(a)$  defined in (\ref{eq:Wntilde}) and $a(-s)$ calculated  by solving the renormalization-group equation (\ref{eq:rge}) iteratively along the circle starting from a given $a(m_\tau^2)$. 

The standard FO expansion of  $\delta^{(0)}$ writes as
\beq\label{eq:del0gen}
\delta_{\rm FO}^{(0)} =\sum_{n\ge 1} d_n \,[a(m_\tau^2)]^n, 
\eeq
where the coefficients $d_n$ are obtained by integrating the $s$-dependent coefficients of  (\ref{eq:hatD}) along the circle.
In order to obtain the improved expansion, we start from the Borel transform $B_{\delta}(u)$ associated to the series (\ref{eq:del0gen}), defined by the relations (\ref{eq:Bdelta}) and (\ref{eq:bndelta}), and expand it in powers of the variable $w$.
In the previous works \cite{Caprini:2009vf, Caprini:2011ya, Abbas:2013usa}, the factorization of the first singularities in this expansion was done  assuming that the nature of these singularities of $B_{\delta}(u)$  and $B_{\wh D}(u)$ is the same. By inserting in (\ref{eq:Laplacedelta0}) the expansion
 \beq\label{eq:Bdelta0wold}
B_{\delta}(u)=\frac{1}{(1+w)^{2\gamma_{1}} (1-w)^{2\gamma_{2}}} \sum_{n\ge 0} {\widetilde \delta}_n\, w^n,
\eeq 
 the improved FO expansion of $\delta^{(0)}$ considered in \cite{Caprini:2009vf, Caprini:2011ya, Abbas:2013usa} had the form
\be\label{eq:del0FOold}
\delta_\text{FO}^{(0)}=\sum_{n\ge 0} {\widetilde \delta}_n \,{\widetilde {\cal W}}_n(a(m_\tau^2)), 
\ee
with $ {\widetilde {\cal W}}_n(a)$  defined in (\ref{eq:Wntilde}).
 However, the relation 
\beq\label{eq:BdeltaBD}
B_\delta(u)=\frac{12}{(1-u)(3-u)(4-u)} \frac{\sin(\pi u)}{\pi u} B_{\wh D}(u),
\eeq
established in \cite{Brown:1992pk, Beneke:2008ad} in the large-$\beta_0$ approximation and shown in \cite{Boito:2020hvu} to hold in general QCD in the $C$-scheme, suggests that  the  singularities  of  $B_\delta(u)$    at $u=-1$ and $u=2$ might be milder than those of $B_{\wh D}(u)$. In the extreme case, making the conjecture that these singularities are weakened by simple zeros as in (\ref{eq:BdeltaBD}), we write
\beq\label{eq:Bdelta0w}
B_{\delta}(u)=\frac{1}{(1+w)^{2(\gamma_{1}-1)} (1-w)^{2(\gamma_{2}-1)}} \sum_{n\ge 0} {\widetilde \delta}_n'\, w^n.
\eeq
By inserting this expansion into the Laplace-Borel integral (\ref{eq:Laplacedelta0}), we obtain the alternative expansion
 \be\label{eq:del0FOnew}
\delta_\text{FO}^{(0)}=\sum_{n\ge 0} {\widetilde \delta}_n'\, {\widetilde {\cal W}}'_n(a(m_\tau^2)),
\ee
where
\be\label{eq:Wntilde'}
\hspace{-0.05cm}{\widetilde {\cal W}}_n'(a)=\frac{1}{\beta_0}{\rm PV} \int\limits_0^\infty\!  \frac{{\rm e}^{-\frac{u}{\beta_0 a}}(\tilde w(u))^n}{(1+\tilde w(u))^{2(\gamma_{1}-1)} (1-\tilde w(u))^{2(\gamma_{2}-1)}}du.
\ee

The good convergence of the expansions of $\delta^{(0)}$ improved by the conformal mapping of the Borel plane has been demonstrated numerically in \cite{Caprini:2009vf, Caprini:2011ya, Abbas:2013usa} on realistic models of the Adler function. The numerical studies have shown also that the CI expansion  (\ref{eq:del0CI}) gives  better results than the FO expansion  (\ref{eq:del0FOold}), based on the assumption that the nature of the first singularities of  $B_{\delta}(u)$ and $ B_{\wh D}(u)$ coincide. But, as discussed above, although the exact nature of the singularities is not known, there are hints that  the  singularities factorized in the expansion (\ref{eq:del0FOold}) are stronger than needed. In order to illustrate the dependence on the factorization, it is instructive to investigate also the extreme FO expansion (\ref{eq:del0FOnew}), where the nature of the first branch points is modified as in the large-$\beta_0$ approximation.

\begin{figure}
\includegraphics[width=6.8cm]{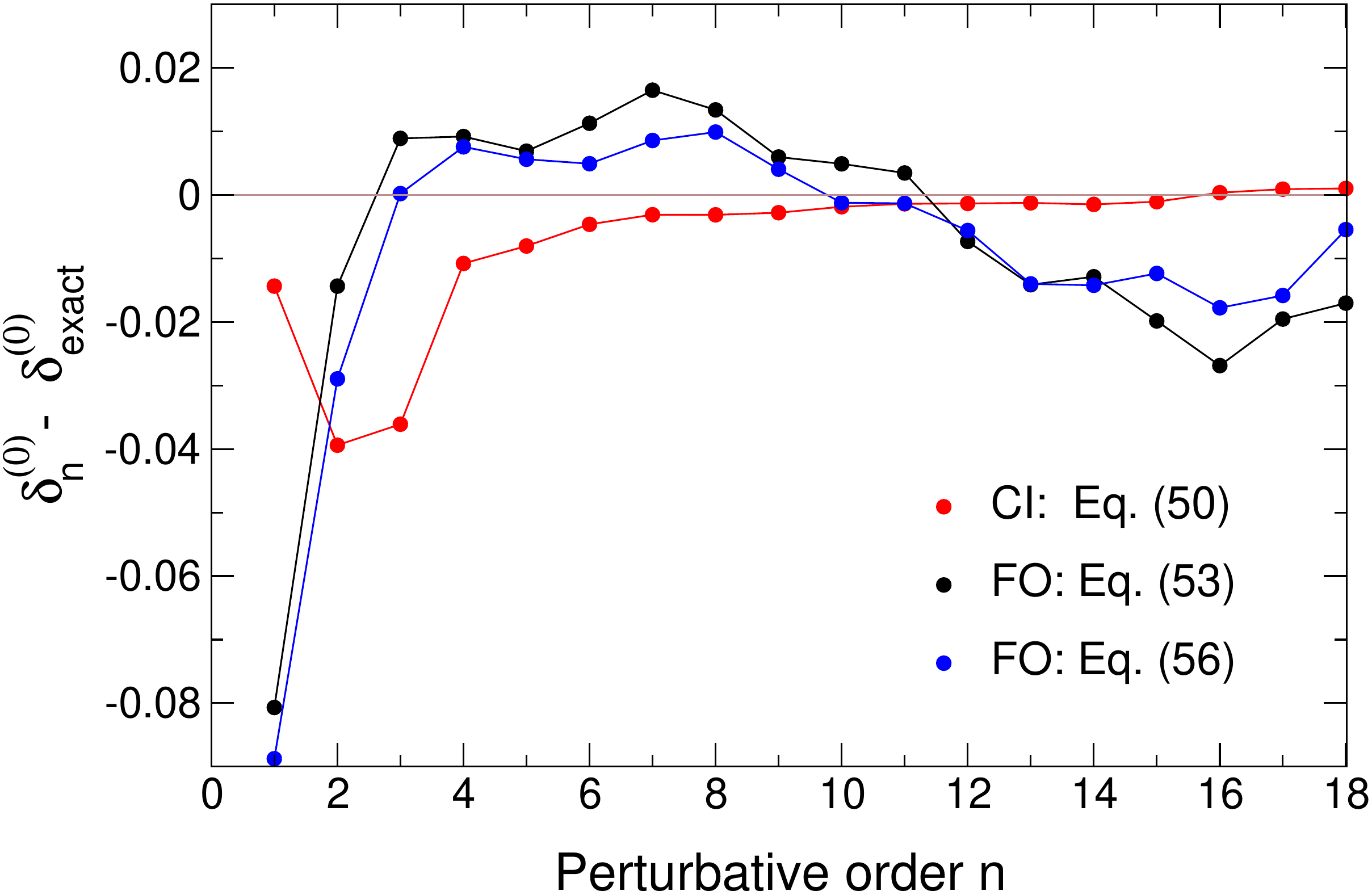}
\caption{Difference between the approximate and the exact values of $\delta^{(0)}$  as a function of the perturbative order $n$ for the reference model proposed in \cite{Beneke:2008ad}.
\label{fig:CIFO}}
\end{figure}

For a numerical test, we consider the ``reference model'' proposed in \cite{Beneke:2008ad}, which gives  for $\alpha_s(m_\tau^2)=0.34$ the exact value  $\delta^{(0)}_\text{exact}=0.2371$, and show in Fig. \ref{fig:CIFO}  the difference between the predictions of the perturbative expansions truncated at order $n$ and the exact value.  We present the results obtained with the CI expansion (\ref{eq:del0CI}) and the FO expansions (\ref{eq:del0FOold}) and (\ref{eq:del0FOnew}). The results for the first two expansions have been reported already in Fig. 2 of  \cite{Caprini:2009vf}.

  Figure \ref{fig:CIFO} shows that all the expansions improved by the conformal mapping of the Borel plane have a tamed behavior at large orders, remarked already in the previous works  \cite{Caprini:2009vf, Caprini:2011ya, Abbas:2013usa}. By contrast, the standard expansions exhibit wild oscillations at large orders (see for instance  Fig. 1 of  \cite{Caprini:2009vf}).  The figure shows also that the improved CI expansion (\ref{eq:del0CI}) converges to the exact value of $\delta^{(0)}$, while the FO expansions exhibit  oscillations around the exact value  up to high orders.  The FO expansion (\ref{eq:del0FOnew}) leads to a slightly better approximation compared to the FO expansion (\ref{eq:del0FOold}), but the improvement is rather modest. 

As discussed in  \cite{Caprini:2009vf, Caprini:2011ya}, the better approximation provided by the improved CI expansion is explained by the fact that it simultaneously  implements renormalization-group invariance and accelerates the convergence of the perturbative series by exploiting the known large-order behavior of the expanded function. By contrast, the improved FO expansions treat only one facet of the problem: they accelerate the convergence of the perturbative series (\ref{eq:del0gen}), but do not cure the poorly convergent expansion (\ref{eq:hatD})  of the Adler function in the complex plane, especially near the 
timelike region\footnote{This fact is clearly illustrated in Figs. 14 and 15  of \cite{Caprini:2009vf} and in Figs. 2-5 of \cite{Caprini:2011ya}, where the values of the Adler function  in the complex $s$ plane, along the contour
$|s|=m_\tau^2$, are shown.}. The more solid theoretical basis and the good convergence properties proved numerically  make the nonpower CI expansion (\ref{eq:del0CI}) the best option for calculating the $\tau$ hadronic width in perturbative QCD. 
 
\section{Strong coupling from $\tau$ hadronic width}\label{sec:alphas}
The expansions improved by the conformal mapping of the Borel plane have been used for the extraction of the strong coupling $\alpha_s(m_\tau^2)$ from the $\tau$ hadronic width in  \cite{Caprini:2009vf, Caprini:2011ya}. In this work we present an update of this determination. The main new ingredient is the conjecture,  formulated and discussed  in this paper, that the nonpower expansions recover nonperturbative features of the expanded function, making unnecessary the addition of the power corrections.

As experimental input, we use the difference,  quoted  in \cite{Bethke:2011tr}, page 25,  between the phenomenological value of $\delta^{(0)}$ and the PC contribution to it,  estimated in \cite{Beneke:2008ad} to be $-7.1\times 10^{-3}$. After  adding back this term we obtain the phenomenological value
\be\label{eq:input}
\delta^{(0)}_{\rm phen}=0.1966 \pm 0.0040_{\rm exp}.
\ee
On the theoretical side,  we use the expansion (\ref{eq:del0CI}) truncated  at $n=5$, with the coefficients $\wt c_n$ given in (\ref{eq:cntilde})
and the expansion functions $ {\widetilde {\cal W}}_n(a)$  defined in (\ref{eq:Wntilde}). As we mentioned above, while the optimal conformal mapping (\ref{eq:w}) is unique, the factorization of the first singularities is not. Therefore, in the assessment of the theoretical uncertainty we accounted also for other possibilities of factorization. 

 The running coupling $a(-s)$ was calculated by solving the renormalization-group equation (\ref{eq:rge}) iteratively along the circle, starting from $s=-m_\tau^2$. 
For completeness, we  investigated also other scales  by setting in (\ref{eq:hatD}) more generally $\mu^2=-\xi s$ with $\xi =1\pm 0.63$ \cite{Davier:2008sk, Pich:2013lsa} and applying to the resulting series the steps leading to the improved expansion. The theoretical expression depends implicitly on the value of $\alpha_s(m_\tau^2)$, which was found numerically from the phenomenological input as
\beq\label{eq:alpha}
\alpha_s(m_\tau^2)=0.314 \pm 0.004_\text{exp}\,\pm 0.003_{c_{5,1}}~^{+0.002}_{-0.001} \,\text{(scale)},
\eeq
where we indicated the separate sources of error. By combining these errors in quadrature and adding a conservative error of 0.001 to account for other ways of softening the first singularities, we obtain
\beq\label{eq:alpha1}
\alpha_s(m_\tau^2)=0.314 \pm 0.006.
\eeq
Compared to the previous determination $0.320 \pm 0.020$ quoted in \cite{Caprini:2011ya}, the difference is due mainly to the conjecture made now on the PC contribution,  which leads to the shift by 0.006 of the central value  and a slight reduction of the error. We note   also that a different value $c_{5,1}=283\pm 283$ was used in  \cite{Caprini:2011ya}, instead of the more precise estimate (\ref{eq:c51}) obtained in \cite{Boito:2018rwt, Caprini:2019kwp}. Moreover, in the calculation of $a(-s)$ along the circle we now used the $\beta$ function to five-loop, derived recently in \cite{Baikov:2016tgj}.
 
Using the standard packages \cite{Herren:2017osy} for running the coupling  and adding an error of 0.0003 due to evolution, we find
\beq\label{eq:alphamZ}
\alpha_s(m_Z^2)=0.1179 \pm 0.0008,
\eeq
which practically coincides with the world average $\alpha_s(m_Z^2)=0.1179 \pm 0.0010$ quoted in the latest version of PDG \cite{Zyla:2020zbs}.

~\vspace{-0.5cm}

\section{Discussion and conclusions}\label{sec:conc}
In perturbative QCD, the expansions truncated at finite orders depend on the renormalization scheme and scale, violating the renormalization-group invariance of the full theory.  Also,  the perturbation series are expected to be divergent, with  coefficients growing factorially at large orders, being at most asymptotic expansions to the exact functions.  These two properties are related: for instance, contrary to na\"{i}ve expectations, the inclusion of additional terms in the expansion  of the $\tau$ hadronic width did not reduce the dependence on the renormalization-group prescription.

 The treatment of the divergent expansions in  perturbative QCD can be related formally to the mathematical concept of hyperasymptotics, which amounts to a sequence of truncated "transseries", each exponentially small in the expansion parameter of the previous one,  which allow the exact function to ``resurge''.  In QCD, the first additional series is associated to the  power corrections in the standard OPE,  which supplement the truncated perturbation series and recover some of the nonperturbative features of the exact function. 

In the present work, we discussed a reformulation of QCD perturbation theory as an expansion in terms of a set of nonpower functions of the strong coupling. These functions are defined through the  analytic continuation in the Borel plane, achieved by  the optimal conformal mapping of this plane. The new expansions have been defined and investigated in \cite{Caprini:1998wg, Caprini:2000js, Caprini:2001mn}, and extensive numerical studies and applications have been performed in the subsequent works \cite{Caprini:2009vf, Caprini:2011ya, Abbas:2012fi, Abbas:2013usa}. In this paper, we reviewed the theoretical properties of these expansions  and argued that they can be viewed as an alternative to the transseries for recapturing nonperturbative features of the exact QCD correlators. 

 For the Adler function, the new expansion is given in  (\ref{eq:cW})  in terms of the expansion functions (\ref{eq:Wn}). An improved version, which exploits also the known nature of the first singularities in the Borel plane is given in (\ref{eq:cWtilde}) and  (\ref{eq:Wntilde}). As discussed in Sec. \ref{sec:prop},  the expansion functions have properties similar to the expanded correlator: they are singular at the origin of the coupling plane and their  perturbative expansions in powers of $\alpha_s$ are divergent series.  The new expansion incorporates therefore nonperturbative features, much like the power corrections in the OPE representation (\ref{eq:OPE}): both contain terms of the form $\exp (-c/\alpha_s)$, singular at  $\alpha_s=0$.  Moreover, while the standard perturbation series is divergent, the new expansions  have a tamed behavior at large orders and may even converge in some conditions. 

Using these properties and theoretical arguments based on the Borel plane,  we formulated the conjecture that the method of conformal mapping can be an alternative to the transeries approach for dealing with the divergent expansions in QCD. This means that the new expansions  (\ref{eq:cW}) or (\ref{eq:cWtilde}) are able to recover nontrivial nonperturbative features of the QCD correlators,  without the need of additional, arbitrary power corrections. 

Two further arguments can be invoked in support of this assumption, as discussed in Sec. \ref{sec:compar}. First, when reexpanded in powers of the coupling, the new expansions, even truncated at finite orders, contain a infinite number of terms. By contrast, the standard OPE contains a truncated perturbation expansion and, as discussed in recent analyses \cite{Bali:2014sja, Ayala:2019uaw, Ayala:2019hkn}, the nonperturbative terms  depend  on the truncation order of this expansion. So, if one may think to add arbitrary power corrections  to the new expansions, their interpretation in terms of condensates will be hard to give. 

The second argument is based on the fact that the new expansion is shown to converge under some conditions (not proved, but expected to be valid in QCD), and the convergence is checked numerically  on models inspired from QCD.  Since there are  no  reasons for adding new terms to
a convergent series, we conclude that there are no mathematical arguments  for supplementing the nonpower expansions by other, arbitrary power corrections.

We note that similar conclusions have been obtained recently in several mathematical works \cite{Costin:2017ziv, Costin:2019xql, Florio:2019hzn}, where the possibility of resurgence from pure perturbation theory,  without additional transseries,  was demonstrated numerically in specific cases where the exact function is known. 

 In QCD, as discussed above, both OPE and the present expansions (\ref{eq:cW}) or (\ref{eq:cWtilde}) account for power corrections, i.e. for singularities  of the form $\exp(-c/\alpha_s)$ at the origin of the coupling plane. Therefore, they do not reproduce the complicated singularity structure at this point of the exact correlator, proved in \cite{tHooft}.  The existence of additional, "duality-violating" contributions has been recently advocated  \cite{Blok:1997hs, Shifman:2000jv, Cata:2005zj, Peris:2016jah, Boito:2017cnp} in order to better approach the physical correlator. These terms, which  decrease exponentially on the Euclidian axis and exhibit an oscillating behavior when analytically continued to the timelike axis, are not easy to parametrize and obscure the determination of the
nonperturbative condensates in the standard OPE. Since the duality-violating contributions  go beyond the power corrections, they are expected to show up also in addition to the new expansions discussed in this paper. A phenomenological investigation of this problem is beyond the scope of this paper and will be considered in a future work. 

Actually, since neither OPE nor the new expansions discussed in this paper are able to describe the hadronic resonances and the unitarity thresholds present in the spectral functions of correlators, they can be confronted to experiment only for ``smeared'' observables, as remarked a long time ago in \cite{Poggio:1975af}. 
Alternatively, integrated observables like the moments of the spectral function have been much used in phenomenological studies,  because they can be expressed as weighted integrals of the Adler function along a contour in the complex plane.  
We discussed in  Sec. \ref{sec:moments}  the improved expansions  based on conformal mapping for the moments, in both CI and FO versions of renormalization-group summation. We also reviewed recent results on the singularities of the FO expansions of the moments in the Borel plane,  which show that in the $\overline{{\rm MS}}$ scheme the nature of the first singularities is not known exactly, although the large-$\beta_0$ approximation may provide a hint. 

 In Sec. \ref{sec:delta0}, we discussed in particular  the improved expansions based on conformal mapping for the $\tau$ hadronic width\footnote{Since the weight (\ref{eq:Wtau}) suppresses the region near the timelike axis, the  duality-violating contributions can be neglected in this case.}.   As seen from  Fig. \ref{fig:CIFO}, these expansions have a tamed behavior at large orders for both CI and FO resummations. The figure shows also that the CI expansion  (\ref{eq:del0CI}) approximates the exact value more precisely than the  FO expansions (\ref{eq:del0FOold}) and  (\ref{eq:del0FOnew}), defined with two extreme assumptions about the nature of the first renormalons.  The better convergence is due to the fact that the  CI expansion  implements  simultaneously the renormalization-group improvement and the acceleration of the perturbative series, while the FO expansions accelerate the convergence of the perturbative series, but do not cure the poorly convergent expansion (\ref{eq:hatD})  near the timelike region.  The conclusion is that the CI expansion  (\ref{eq:del0CI}) has a more solid theoretical basis and is the best option for physical applications. 

Finally, as an illustration of our approach, we presented in Sec. \ref{sec:alphas}  an updated determination of the strong coupling from $\tau$ hadronic width. The reformulation of perturbative QCD  by the conformal mapping of the Borel plane and the conjecture about the PC contribution made in this work  lead to a  reduction of the central value of $\alpha_s(m_\tau^2)$  and a slightly smaller uncertainty. The  precision is further improved  by using recent estimates of the six-loop perturbative coefficient of the Adler function.  Our prediction is given in (\ref{eq:alpha1}) and implies for $\alpha_s(m_Z^2)$ the value (\ref{eq:alphamZ}), practically identical to the present world average quoted in \cite{Zyla:2020zbs}.

\subsection*{Acknowledgments}
I thank D. Boito for useful discussions. This work was supported by the Ministry of Education and Research, Romania, Contract PN 19060101/2019-2022.


\end{document}